\documentclass[sigconf]{acmart}
\usepackage{xspace}
\usepackage{graphicx}
\usepackage{booktabs}
\usepackage{array}
\usepackage{colortbl}
\usepackage{subcaption}
\usepackage{enumitem}
\usepackage{float}
\setcopyright{none}
\renewcommand\footnotetextcopyrightpermission[1]{}
\newcommand{\tightcaption}[1]{\caption{#1}}
\newcommand{\colortablerules}{\setlength{\aboverulesep}{0pt}\setlength{\belowrulesep}{0pt}\renewcommand{\arraystretch}{1.15}\color{black}}
\newcommand{\llmdominant}{MGT-dominant\xspace}
\newcommand{\sysname}{DeGenTWeb\xspace}
\newcommand{\parab}[1]{\noindent{\bf #1}\ \ignorespaces}

\newcommand{\squishlist}{\begin{itemize}[leftmargin=*]}
\newcommand{\squishend}{\end{itemize}}

\begin{document}

\title{\sysname: A First Look at \llmdominant Websites}

\author{Sichang Steven He}
\affiliation{\institution{University of Southern California}\city{Los Angeles}\country{USA}}
\orcid{0009-0009-2565-1508}

\author{Calvin Ardi}
\affiliation{\institution{Independent Researcher}\country{USA}}

\author{Ramesh Govindan}
\affiliation{\institution{University of Southern California}\city{Los Angeles}\country{USA}}
\orcid{0000-0001-8311-8853}

\author{Harsha V. Madhyastha}
\affiliation{\institution{University of Southern California}\city{Los Angeles}\country{USA}}
\orcid{0009-0000-3921-5680}

\begin{abstract}
    Many news reports have claimed that machine-generated text (MGT) from large
language models (LLMs) is taking over the web.
However, these claims are typically not based on a representative sample of
the web, and the methodology underlying them is often opaque.
Consequently, we lack an understanding of the true prevalence and
characteristics of MGT on the web.

We describe \sysname{}, which systematically identifies \llmdominant websites:
sites whose informational prose content is mostly MGT, with
little human input.
We show how to adapt detectors of \emph{MGT} for use on \emph{web pages}, and
how to aggregate detection results from multiple pages on a site for accurate
\emph{site-level} categorization.
\sysname mislabels only 0.12\% of human-authored sites as \llmdominant.
We find that 20.9\% of the
sites archived by Common Crawl between January and July 2026 and 13.6\% of the
sites linked from Bing's search results in 2025 are \llmdominant.
These figures are lower bounds on the prevalence of \llmdominant{} sites.
We also find that the fraction of \llmdominant sites is growing over time, and that many such sites share the same owner.
We will release all of our code and data after publication.

 \end{abstract}

\maketitle

\section{Introduction}\label{sec:intro}

Over three decades ago, Sir Tim Berners-Lee created the web to
enable scientists to seamlessly share information with each other.
Since then, the web has evolved to become arguably the largest distributed
repository of human knowledge in history.
Billions of websites provide access to a wide range of information, which we
draw upon either directly via web searches or indirectly via the responses we
receive from AI-powered chatbots.

However, today, the extent to which users can trust the information they
discover on the web is increasingly under threat.
This is because our assumption that the vast majority of web content has been
authored by humans is being violated.
In the current age of AI, it is natural for users to use LLMs to polish the
prose they author.
But, going beyond that, it is now also possible for a user to create an
\emph{entire website} with many pages simply based on a short textual prompt.
For instance, given the title of a blog and a one-sentence summary of every
blog post on it, state-of-the-art (SoTA) AI website
builders~\cite[e.g.,][]{wix,b12,hostinger,framer,durable} can create all pages on the site.

In this paper, we focus specifically on sites whose pages primarily contain
informational prose.
Examples include sites which offer news articles, how-to guides and tutorials,
descriptions of products and services, educational materials, technical
documentation, etc.
Problems with MGT, such as hallucination and plagiarism, are of particular
concern on such sites because any user relies on their content to learn and
make decisions.
For example, malicious news and commentary sites have been known to use MGT to
spread election-related disinformation~\cite{nyt2026propaganda}.
We refer to sites where most, if not all, of the site's textual content is
informational prose which has been generated using LLMs (i.e., is MGT) as
\emph{\llmdominant}.

This observation that \llmdominant websites are on the rise is not new.
Several
studies~\cite{sadeghi2025tracking,graphite2025more,ahrefs2025newpages,pew2026internet,russell2026token,dolezal2026impact}
have attempted to quantify the prevalence of such sites.
But all of these studies suffer from one of two fundamental drawbacks.
Either the methodology used is opaque, e.g., many studies use proprietary MGT
detectors, whose accuracy is unknown.
Or the studies are conducted on a set of sites which are not representative of
the web at large.

To fill this void, we present \sysname,\footnote{Short for either Detect
Generated Text on the Web, or Degenerate Web.} the first systematic
methodology to identify \llmdominant websites, and apply it to a large swath
of the web.
The primary challenge is that even the best MGT detectors are unreliable when
configured to minimize the odds of falsely flagging human-authored
text~\cite{dugan2024raid}.
Moreover, classifiers of \emph{chunks of text} cannot readily be applied to
\emph{web pages}, wherein multiple snippets of text are separated by HTML
markup and boilerplate content is repeated across multiple pages on a site.
In addressing these challenges, we make four contributions.

First, we show that the key to accurately identifying the use of MGT on the web
is to categorize at the granularity of a \emph{website} and to carefully
select which subset of pages to analyze on each site.
Though an MGT detector can be inaccurate on some pages, \sysname can
accurately categorize a site by aggregating the detector's scores for multiple
pages on the site.
We perform this aggregation only across pages which, after we identify and
exclude boilerplate content, contain a sufficient amount of prose for the
detector to be reliable.

Second, to scientifically validate \sysname, we assemble a new ground truth
dataset of 1,256 \llmdominant and 10,328
non-\llmdominant sites.
We find that sampling 15 pages per site suffices to achieve accurate
classification.
We show that, via judicious filtering of pages and aggregation of the MGT
detector's scores, we are able to achieve low false positive (0.12\%) and
false negative ($\sim$2\%) rates on our ground truth dataset.

Third, we use \sysname to classify 117k sites archived by Common Crawl and
18k sites found in Bing's results for 10k how-to search queries.
We find that the fraction of \llmdominant sites in Common Crawl's data
has been steadily growing over the last few years: up from
2.0\% in late 2022 to
20.9\%
in early 2026.
We also find that \llmdominant sites are highly prevalent in Bing's search
results: for about two-thirds of our queries, the top 20 results point to at
least one \llmdominant site.
These numbers are, however, lower bounds on the prevalence of \llmdominant sites.
This is because, in the interest of scalability and transparency, we use
inexpensive open source MGT detectors, even though they fail to identify text
generated by the newest LLMs~\cite{yu2025evobench,wu2024detectrl}.
We find that \sysname would be able to identify these sites if paired with a
proprietary detector, but that would cost over $300\times$ more.

Lastly, we uncover the influence of financial incentives.
Compared to other sites, \llmdominant sites more often sell goods and services
and are less often personal or organizational.
We find that they are also less likely to attempt user retention (e.g., via
subscriptions) and instead appear to monetize from one-off visits.

\section{Datasets}\label{sec:datasets}

\autoref{tab:datasets} introduces the 6 datasets we use in this paper,
totaling 147k sites\footnote{We consider every fully qualified domain name as
a unique website.} and 2.2M pages ($\sim$15 pages per site).
We use the first four datasets to train and test \sysname, and the last two to
quantify the prevalence of \llmdominant sites in the wild.

\begin{table}[t]
    \centering
    \tightcaption{The datasets we assemble for our study.}
    \label{tab:datasets}
    \setlength{\tabcolsep}{3pt}
    \colortablerules
	{\small
    \begin{tabular}{@{}lrp{0.6\linewidth}@{}}
    \toprule
    Dataset & \#Sites & Source of data \\
    \midrule
    CC2014 & 10,000 & Common Crawl archive (Oct. 2014) \\
    Body-swap & 1,172 & Page description $\rightarrow$ LLM $\rightarrow$ new page body \\
    Polished & 328 & Page body $\rightarrow$ LLM $\rightarrow$ polished body \\
    Wix/B12 & 45/39 & Commercial AI website builders (2025) \\
    New CC & 117,069 & Common Crawl archive (2020--Jul. 2026) \\
    Search & 18,169 & How-to query $\rightarrow$ Bing $\rightarrow$ 20 results (2025) \\
    \bottomrule
    \end{tabular}
	}
\end{table}

\subsection{Building an \llmdominant baseline dataset}\label{sec:baseline-dataset}

To train and test \sysname's website classifier, we need a ground-truth
dataset where we know whether each site is \llmdominant.
There are many benchmark datasets for evaluating MGT detectors, but no such
dataset exists for web pages or websites.
To fill this gap, we compiled a dataset of
11,584 websites, totaling
174k pages in all, as follows.

\parab{Curating non-\llmdominant sites (``CC2014'').}\label{sec:cc-sampling}
We used HTMLs archived by Common Crawl (CC)~\cite{commoncrawl2025}, which is
the largest publicly available sample of the
web~\cite{thompson2024improved,baack2024critical}.
Following existing common practices in curating human-authored
text~\cite{mitchell2023detectgpt,verma-etal-2024-ghostbuster,wu2024detectrl,dugan2024raid},
we sampled pages archived in 2014, before modern generative language models
were prevalent in the public~\cite{greff2017lstm,karpathy2015visualizing}.
To randomly sample sites archived by CC, we downloaded every
32\textsuperscript{nd} of its October 2014 index files.
From those, we selected records of type text/html with status code 2xx in which
CC detects English text.
For uniformly random yet reproducible sampling, we selected subdomains based on
the BLAKE3 cryptographic hash~\cite{o2021one} of every subdomain.
For each selected subdomain, we uniformly sampled up to 640 pages, again based
on BLAKE3 hash.
Among the sampled sites with at least 15 qualifying pages each, we selected 10k;
we describe later in \S\ref{sec:pipeline} how we filter for pages with
sufficient prose-like text.

\parab{Creating \llmdominant{} sites (``Body-swap'').}
We lack reliable information as to which sites on the web are \llmdominant.
Therefore, we generated such sites ourselves.
For every sampled page of each of 1,172
\emph{CC2014} sites,\footnote{The number of sites is constrained by our budget
of \$300 on OpenRouter.} we swapped the main body of its HTML with MGT,
leaving the rest untouched.
We first used an LLM to briefly describe the original page.
Then, in a fresh context with only that description, we used the LLM to
generate the HTML with the new body.
Each site used one of eight LLMs in turn: Claude Haiku 4.5, Sonnet 4, Sonnet
4.6, GPT-3.5 Turbo, GPT-4, GPT-OSS-120B, Llama 3.3 70B, and Mixtral 8x22B; we
consider more recent models later in \S\ref{sec:newer-wix}.
Each \emph{Body-swap} page retains the original page's metadata, layout, and
boilerplate like sidebar and footer.

\parab{Creating non-\llmdominant sites (``Polished'').}
We do not wish to classify a site as \llmdominant if its pages have mixed
human and LLM authorship, e.g., if a human writes the page's text first and
then uses an LLM to polish it.
Hence, just as we created the body-swap sites, we created
328 \emph{Polished} sites.
The only difference is that we directly gave LLMs the text from the original
page and asked them to rewrite it (see \S\ref{sec:prompt} for prompts).

\parab{Building out-of-distribution \llmdominant{} sites (``Wix/B12'').}
To evaluate \sysname on \llmdominant sites that are unlike our
\emph{Body-swap} dataset and better represent real-world websites, we used AI
website builders from Wix.com and B12.io.
Using sites of random Russell 2000 companies~\cite{russell20002025} as seeds,
we asked ChatGPT to a) describe each website in one paragraph, b) suggest a
name for a similar website, and c) write descriptions for 30 blog posts (see
\S\ref{sec:prompt} for the prompt used).
We manually input each generated website's name and description into the web
interface of Wix.com's AI website builder to generate the home page and
boilerplate pages.
We then input blog post descriptions to generate them one by one, randomizing
each option: suggested blog title selection, blog length, style, and number of
images.
Similarly, we used B12.io to generate sites seeded from IndieWeb personal
blogs~\cite{indieweb2025} to produce a distribution distinct from the Wix
sites.

In April 2025, we used this process to generate 45 Wix
sites and 39 B12 sites.
We also tried AI website builders Hostinger, Framer, 10Web, and
Durable~\cite{hostinger,framer,tenweb,durable}.
But, their free trials were inadequate and their paid plans were beyond our
means (\$20 to \$120 per site).

\subsection{Sampling sites in the wild}\label{sec:wild-datasets}

To estimate the prevalence of \llmdominant sites across the web, we studied
sites from two sources: Common Crawl for a broad sample of the web, and Bing
search results that capture what people searching for information see.

\parab{New CC Dataset.}
First, we randomly selected 1,297,346 of the
sites archived by Common Crawl (CC)~\cite{commoncrawl2025}.
In the same way we curated the \emph{CC2014} sites in
\S\ref{sec:baseline-dataset}, we uniformly sampled sites archived between
January 2020 and July 2026, and then filtered out pages with insufficient
non-boilerplate text (\autoref{fig:page-filter-breakdown}).
We were left with 117,069 sites with at least 15
pages each.

\parab{Search Dataset.}
To study sites encountered by users, we issued 10k search queries to Bing
between February and August 2025.\footnote{Unlike Google, results with Bing's
search API matched its web search.}
We crawled the sites which appeared in the search results.
To mimic common how-to searches, each of our search queries used the title
from a random wikiHow article~\cite{wikihowzim2023}.
The top 20 results for these queries collectively point to
59,046 sites.
For each site, we sampled pages randomly either from the site's sitemap or from
Wayback Machine's index for the site.
For 18,169 sites, we were able to crawl 15
pages which passed our filtering criteria.

\section{Detecting \llmdominant Websites}\label{sec:method}

\begin{figure*}[t]
    \centering
    \includegraphics[width=0.9\linewidth]{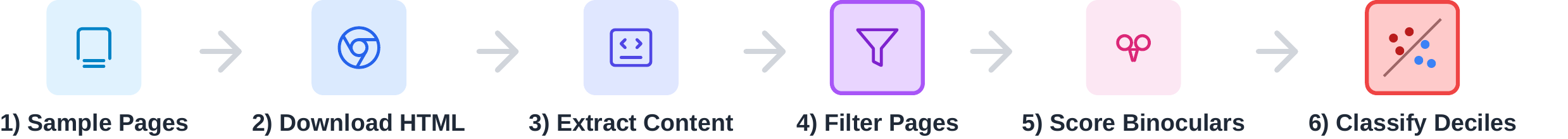}
    \tightcaption{\sysname pipeline for detecting \llmdominant websites.}

    \label{fig:pipeline-overview}
\end{figure*}

To identify \llmdominant websites, we leverage state-of-the-art detectors of
MGT and address their limitations.
We use Binoculars~\cite{hans2024spotting}, a zero-shot detector which
does not need to be rebuilt in order to detect text from every new LLM.
Given a chunk of text, Binoculars outputs a score; the lower the score, the
greater the likelihood the text was generated by an LLM.
Binoculars has been shown to be accurate across many topical domains and
source LLMs, is robust to many adversarial attacks,
and is the most accurate at low false positive rates
(FPR)~\cite{hans2024spotting,dugan2024raid,bevendorff2025two,koike2025machine};
among the 15 MGT detectors we tried, we too find Binoculars consistently gives
the lowest FPR (\S\ref{sec:eval-detectors}).
We also find Binoculars to be accurate in differentiating MGT from
LLM-assisted text (\S\ref{sec:eval-binoculars}), consistent with prior
studies~\cite{li-etal-2024-mage}.

\subsection{Challenges and Approach}

Identifying MGT on the web is, however, not as simple as executing Binoculars
on every page.
First, not all text on web pages is in the form of continuous prose, which MGT
detectors expect as input; e.g., the text on a page may belong to a list of
links or an interactive interface.
Second, MGT detectors are less accurate when executed on short text, e.g.,
Binoculars' false negative rate (FNR) rises from 4\% on text with 256 tokens
to 20\% on text with 128 tokens on one
dataset~\cite{hans2024spotting}.
Third, Binoculars offers a poor tradeoff between false positive and false
negative rates; lowering FPR from 5\% to 1\% increases FNR from
20\% to 27\% on RAID~\cite{dugan2024raid} and from 33\% to
42\% in~\cite{tufts2025practical}.
A low FPR is essential in the context of the web where, at least for now,
human-authored prose dominates.
For instance, if 5\% of sites were \llmdominant, a 1\% FPR would mean that,
even with no false negatives, roughly one-sixth of the sites classified as
\llmdominant would be incorrectly flagged as such.

We address these limitations based on two insights.
\squishlist
\item \textbf{Aggregate page signals for robust site-level detection.}
We observe that if the same source---LLM or human---produces many pages on a site, Binoculars' correct classifications eventually outnumber its misclassifications.
Thus, for high accuracy, we opt to aggregate signals across multiple pages of the same site for a \emph{site-level} classification.
While some sites may have a mix of MGT and human-authored pages, our approach helps alert users about the quality of content on any site whose pages are consistently classified as MGT.

\item \textbf{Focus on pages with enough non-boilerplate content.}
We must carefully consider which subset of pages to analyze on any site and what portion of the content on each of these pages to consider.
On the one hand, many websites include boilerplate content which is present on all of the site's pages.
Hence, instead of feeding the text on any web page as is into Binoculars, we must identify and exclude such boilerplate text.
On the other hand, we only consider pages which include sufficient content for Binoculars' classification to be reliable.
\squishend

Below, we describe how we realize these insights in \sysname.
Then, in \S\ref{sec:eval-pipeline}, we validate our approach.

\subsection{\sysname Detection Pipeline}\label{sec:pipeline}

As \autoref{fig:pipeline-overview} illustrates, to categorize a website, we
must first crawl multiple pages from the site.
For CC sites, we directly use their archived HTMLs.
For others, we load each sampled page in a Chromium-based headless browser and
save the page's final HTML.
We then use Trafilatura~\cite{barbaresi2021trafilatura}, a SoTA web content
extraction tool~\cite{bevendorff2023empirical}, to extract the main textual
content from the HTML.
It returns the content on a page that one would expect to see when
the page is loaded in a browser's Reader mode, i.e., no boilerplate,
recommendations, etc.

\parab{Page filtering.}
Because our focus is on informational prose and Binoculars is less accurate on
shorter text~\cite{hans2024spotting}, we only consider pages which have
English content with 200 or more Falcon-7B tokens (used by Binoculars).
Next, we discard pages which do not pass Dolma's quality
filter~\cite{soldaini2024dolma}.
Designed for LLM training data curation, Dolma's quality filter suits our
exact needs to filter out non-prose text~\cite{soldaini2024dolma}.
Following prior common practices~\cite{wenzek2020ccnet,raffel2020exploring,rae2021gopher},
Dolma drops lines that do not end in punctuation, then rejects texts with
heavy $n$-gram or line repetition, few alphabetic words, too few common
English words, or too many lines that start with bullet points or end with
ellipses, as detailed in~\S\ref{sec:dolma}.
Lastly, we exclude pages on which more than half the page content is found on
other pages on the same site, so that most of the text on each page we judge
is unique to that site; we divide each page's content into chunks using Rabin
fingerprinting~\cite{manber1994finding} and detect exact duplicates of chunks
of text.

\begin{table}[b]
    \centering
    \tightcaption{Pages left (\%) after each \sysname filter, and sites left,
	in the datasets from CC and Bing searches. Darker is more filtered out.}
    \label{fig:page-filter-breakdown}
    \colortablerules
    \begin{tabular}{lccc}
\toprule
 & CC2014 & Search & New CC \\
\midrule
All pages & 2.54M & 3.34M & 50.31M \\
Has English text & \cellcolor[rgb]{0.914,0.542,0.249}62.2\% & \cellcolor[rgb]{0.961,0.819,0.546}84.7\% & \cellcolor[rgb]{0.980,0.900,0.737}92.1\% \\
At least 200 tokens & \cellcolor[rgb]{0.461,0.254,0.574}\textcolor{white}{23.2\%} & \cellcolor[rgb]{0.803,0.423,0.324}\textcolor{white}{49.9\%} & \cellcolor[rgb]{0.585,0.296,0.494}\textcolor{white}{32.0\%} \\
Passes Dolma filters & \cellcolor[rgb]{0.397,0.230,0.575}\textcolor{white}{18.4\%} & \cellcolor[rgb]{0.743,0.372,0.379}\textcolor{white}{44.3\%} & \cellcolor[rgb]{0.516,0.273,0.538}\textcolor{white}{27.1\%} \\
At most 50\% duplicated & \cellcolor[rgb]{0.360,0.211,0.520}\textcolor{white}{15.3\%} & \cellcolor[rgb]{0.692,0.331,0.424}\textcolor{white}{39.6\%} & \cellcolor[rgb]{0.426,0.243,0.597}\textcolor{white}{20.7\%} \\
\cmidrule(lr){1-4}
All sites & 32.1K & 44.1K & 356.6K \\
Sites with 15 pages & 10.0K & 18.2K & 117.1K \\
\bottomrule
\end{tabular}

\end{table}

\autoref{fig:page-filter-breakdown} shows the pages our filters retain in
each of the datasets that we collect either from CC or Bing's search queries.
Pages are filtered out mainly because they have no English body text or
are too short.
In particular, because Common Crawl does not crawl pages with a headless
browser, pages which are dynamically rendered have limited text in CC
datasets.
Moreover, we find that CC's classification of English pages is not always
accurate.

\parab{Score aggregation.}
We execute Binoculars on the extracted text for every page that remains after filtering.
For every site, we then compute the 10\textsuperscript{th},
20\textsuperscript{th}, $\ldots$, 90\textsuperscript{th} percentiles of the
per-page Binoculars scores as a \emph{feature vector} for the site.
We input this 9-dimensional vector into a linear support vector machine (SVM;
trained in~\S\ref{sec:eval-pipeline}) which outputs whether the site is
\llmdominant.
We classify a site only when 15 pages remain after filtering; otherwise,
\sysname returns no site label.
We investigate our threshold choices in~\S\ref{sec:eval-pipeline}.

\section{Evaluating \sysname}\label{sec:eval-pipeline}

We evaluate \sysname along several dimensions.
First, we evaluate its accuracy on our datasets where we know the
ground-truth, including in-distribution~(ID) and out-of-distribution~(OOD)
sites.
Second, we validate the need for and sufficiency of our approach via
ablations, including replacing Binoculars with other MGT detectors.
Lastly, we test \sysname on sites generated with newer LLMs and discuss
accuracy versus cost tradeoffs.

We perform 30 trials in each experiment.
In each trial, we randomly choose half of the \emph{Body-swap}
(586) and half of the \emph{CC2014}
(5,000) sites to train \sysname's SVM.
We use each site's 9-decile page score feature vector (\S\ref{sec:pipeline})
and ground-truth label (\llmdominant or not) to train the SVM.
We then perform in-distribution (ID) testing on the remaining \emph{Body-swap}
and \emph{CC2014} sites as well as sites in our \emph{Polished} dataset, and
out-of-distribution (OOD) testing on the sites we generate using Wix and B12.
We see little difference in our ID results even when we attempt to maximize
the difference between the training and test sets, i.e., for every
\emph{Body-swap} site used to train, the CC site from which it is derived is
also in the training set.
This is because our SVM has 10 trainable parameters, which cannot overfit on
the hundreds of samples in the training data.
We report the false positive rate~(FPR)---what fraction of non-\llmdominant
sites we misidentify as \llmdominant---and the false negative
rate~(FNR)---what fraction of \llmdominant sites we miss.
We focus on the 90th percentile~(P90) of FPR and FNR on the test sets across
all trials.

\subsection{Accuracy with Binoculars}\label{sec:eval-binoculars}

\begin{figure}[t]
    \centering
    \includegraphics{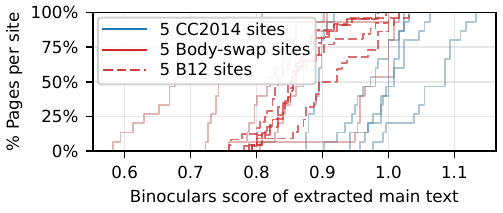}
    \tightcaption{Distribution of per-page Binoculars scores for
    select ground-truth sites. Red is \llmdominant.}

    \label{fig:baseline-svm-cdf}
\end{figure}

\parab{How accurate is \sysname?}\label{sec:accuracy-eval}
Across trials, \sysname achieves a P90 FPR of
0.12\% on ID \emph{CC2014} sites, and a P90 FNR
of 1.5\% on ID \emph{Body-swap} sites.
For the OOD sites generated using Wix and B12, \sysname still maintains a low
P90 FNR: 2.2\% on the 45 Wix
sites and 2.6\% on the B12 sites.
Though the FNR is higher OOD than ID, it is unlikely that overfitting is the
cause, for the reason mentioned earlier.
Of the 328 Polished sites, \sysname only misclassifies
1 to 3 as \llmdominant
in each trial,
resulting in a 0.61\% P90 FPR.

We inspected the 20 CC2014 sites that \sysname
classifies as \llmdominant.
We find two dominant causes for these false positives.
First, Binoculars, and MGT detectors in general, are prone to misclassifying
common text or sentence patterns that they memorized during
training~\cite{hans2024spotting} as the output of LLMs.
For example, \sysname classifies \url{htmlbible.com} as \llmdominant because
that site hosts the Bible, which is often included in the data used to train LLMs.

Second, on sites like \url{shopping1.hp.com}, the body extracted by
Trafilatura retained boilerplate text such as the disclaimer footers.

\begin{figure}[t]
    \centering
    \includegraphics[width=\linewidth]{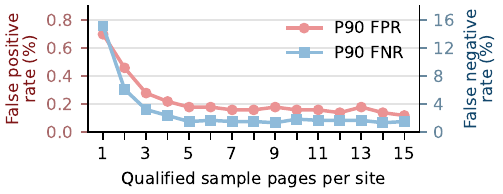}
	\tightcaption{P90 FPR and FNR of \sysname when sampling varying numbers of
	pages per site for the SVM.}

    \label{fig:pages-per-site}
\end{figure}

\parab{Is Binoculars by itself sufficient?}
Across select \emph{CC2014}, \emph{Body-swap}, and B12 sites,
\autoref{fig:baseline-svm-cdf} shows that the range of Binoculars scores
overlaps across \llmdominant and non-\llmdominant sites.
As a result, if we solely relied on classifying pages based on whether
their Binoculars scores are below a threshold, we can correctly classify
the text on only 87.0\% of the pages on
\llmdominant sites as MGT.
If we trained \sysname's SVM using a single ($N=1$) page per site, we get a
low P90 FPR of 0.70\%, but a high P90 FNR
of 15.2\%.
As we increase $N$ in \autoref{fig:pages-per-site}, P90 FPR and FNR fall
steeply until $N=5$, then oscillate and stabilize at
0.12\% and
1.5\% when $N=15$.

\begin{table}[t]
    \centering
    \tightcaption{\sysname's site classification error rates as page filters cumulate.}
    \label{fig:filtering-ablation}
    \colortablerules
    \begin{tabular}{lcc}
\toprule
Filtering condition & P90 FPR \% & P90 FNR \% \\
\midrule
English & \cellcolor[rgb]{0.996,0.966,0.894}1.7\phantom{0} & \cellcolor[rgb]{0.965,0.837,0.589}13.7 \\
+ 200 tokens & \cellcolor[rgb]{0.998,0.976,0.917}0.82 & \cellcolor[rgb]{0.985,0.920,0.785}\phantom{0}6.0 \\
+ Dolma & \cellcolor[rgb]{0.999,0.982,0.932}0.22 & \cellcolor[rgb]{0.997,0.970,0.903}\phantom{0}1.4 \\
+ Deduplication & \cellcolor[rgb]{1.000,0.983,0.935}0.12 & \cellcolor[rgb]{0.996,0.968,0.899}\phantom{0}1.5 \\
\bottomrule
\end{tabular}

\end{table}

\parab{Is filtering of pages necessary?}
We study our choices of the various steps and associated thresholds we use in
filtering pages.
\autoref{fig:filtering-ablation} shows that each of the measures we use for
filtering pages helps decrease FPR and FNR, from
1.7/13.7\%
when allowing all English pages to
0.12/1.5\%
after adding the final deduplication step.
We experiment with different thresholds for the minimum fraction of
non-duplicate content and the minimum number of tokens that a page must
contain.
We varied the former from 10\% to 100\% in steps of 10\%, and the latter from
50 to 400 in steps of 50.
None of these settings lowers P90 FPR below
0.12\%; most increase it.
Thus, we conclude that choosing different thresholds would not substantially
change our results.

\begin{figure}[t]
    \centering
    \includegraphics{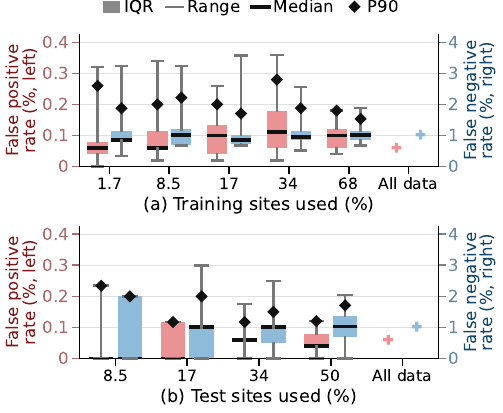}
	\tightcaption{\sysname's error rates when varying the size of either (a)
	the training set or (b) the test set.}

    \label{fig:svm-size-sensitivity}
    \label{fig:svm-training-size}
    \label{fig:svm-test-size-sensitivity}
\end{figure}

\parab{Is our ground-truth dataset large enough?}\label{sec:baseline-size}
We previously measured \sysname's accuracy by randomly splitting the
\emph{CC2014} and \emph{Body-swap} sites in each trial into two equal
partitions comprising the training and test set of sites.
Now, we repeat the same experiment, wherein we continue to test on half of
these sites, but use a random sample of the remaining half to train \sysname's
SVM.
We see little change in FPR and FNR as the training set grows
(\autoref{fig:svm-size-sensitivity}a): P90 FPR is at most
0.28\% at every size, and P90 FNR is
1.9\% with 10 training \emph{Body-swap}
sites and 1.5\% with 400.
We also experiment with the full training set, and subsample the test set.
\autoref{fig:svm-test-size-sensitivity}b shows that neither P90 FPR nor P90
FNR increases as the test set grows.
Thus, we do not expect a significant change in \sysname's error rate with a
larger training or test dataset.

\subsection{Other MGT detectors}\label{sec:eval-detectors}

\parab{How does the choice of MGT detector affect
accuracy?}\label{sec:detector-ablation}
Our results thus far were based on \sysname's SVM using a vector of scores
computed using Binoculars.
We now experiment with several other MGT detectors:
Fast-DetectGPT~\cite{bao2024fast}, L2D~\cite{zhou2026learn},
FastNPR~\cite{sakurai2024impact}, DetectLLM-LRR~\cite{su2023detectllm},
RAIDAR~\cite{mao2024raidar}, BART\-Score~\cite{zhu2023beat},
Entropy~\cite{mitchell2023detectgpt}, IRM~\cite{liu2025implicit},
RoBERTa~\cite{solaiman2019release}, and RADAR~\cite{hu2023radar}.
We also test Log Rank, Log Probability, and
Rank~\cite{mitchell2023detectgpt}, Max Probability~\cite{sakurai2024impact}.
\autoref{tab:detector-ablation} shows that all other options offer comparable
or worse results.
We also tried augmenting Binoculars with 4 other detectors, as well as all 15
detectors together.
Each $k$-detector combination uses all 9 decile scores per detector to form a
$9k$-dimensional feature vector.
Since FPR and FNR depend on the threshold chosen, we show the standard AUROC
for page-level accuracy.
The only combinations that clearly beat Binoculars contain L2D, a
rewrite-based detector which needs auto\-regression and costs $\sim$6$\times$
as much as Binoculars in terms of GPU expenses in our tests.

\begin{table}[t]
    \centering
    \tightcaption{Page-level AUROC (\%), and \sysname's site-level
	error rates (\%) with different MGT detectors. Darker is worse.}
    \label{tab:detector-ablation}
    \setlength{\tabcolsep}{3pt}
    \colortablerules
    \begin{tabular}{lccc}
\toprule
Detector & Page AUROC & P90 FPR & P90 FNR \\
\midrule
Fast-DetectGPT & \cellcolor[rgb]{0.996,0.967,0.896}98.4 & \cellcolor[rgb]{1.000,0.982,0.933}0.18\phantom{0} & \cellcolor[rgb]{0.997,0.970,0.903}\phantom{0}1.4\phantom{0} \\
Binoculars & \cellcolor[rgb]{0.996,0.966,0.894}98.3 & \cellcolor[rgb]{1.000,0.983,0.935}0.12\phantom{0} & \cellcolor[rgb]{0.996,0.968,0.899}\phantom{0}1.5\phantom{0} \\
L2D & \cellcolor[rgb]{0.994,0.961,0.881}97.8 & \cellcolor[rgb]{0.999,0.982,0.932}0.22\phantom{0} & \cellcolor[rgb]{0.993,0.955,0.868}\phantom{0}2.7\phantom{0} \\
FastNPR & \cellcolor[rgb]{0.981,0.906,0.751}92.7 & \cellcolor[rgb]{0.998,0.976,0.917}0.80\phantom{0} & \cellcolor[rgb]{0.967,0.844,0.607}13.0\phantom{0} \\
DetectLLM-LRR & \cellcolor[rgb]{0.977,0.885,0.703}90.8 & \cellcolor[rgb]{0.995,0.963,0.887}2.0\phantom{00} & \cellcolor[rgb]{0.918,0.567,0.269}35.8\phantom{0} \\
RAIDAR & \cellcolor[rgb]{0.971,0.861,0.646}88.6 & \cellcolor[rgb]{0.998,0.975,0.915}0.88\phantom{0} & \cellcolor[rgb]{0.957,0.800,0.503}17.1\phantom{0} \\
BARTScore & \cellcolor[rgb]{0.970,0.858,0.640}88.3 & \cellcolor[rgb]{0.995,0.963,0.887}2.0\phantom{00} & \cellcolor[rgb]{0.927,0.625,0.315}31.2\phantom{0} \\
Entropy & \cellcolor[rgb]{0.963,0.826,0.564}85.4 & \cellcolor[rgb]{0.994,0.959,0.878}2.3\phantom{00} & \cellcolor[rgb]{0.753,0.380,0.369}\textcolor{white}{54.8\phantom{0}} \\
IRM & \cellcolor[rgb]{0.940,0.707,0.379}75.2 & \cellcolor[rgb]{0.999,0.982,0.932}0.24\phantom{0} & \cellcolor[rgb]{0.673,0.325,0.436}\textcolor{white}{61.8\phantom{0}} \\
OpenAI RoBERTa & \cellcolor[rgb]{0.924,0.608,0.301}67.4 & \cellcolor[rgb]{0.998,0.975,0.916}0.84\phantom{0} & \cellcolor[rgb]{0.579,0.294,0.497}\textcolor{white}{68.4\phantom{0}} \\
RADAR & \cellcolor[rgb]{0.923,0.602,0.297}67.0 & \cellcolor[rgb]{0.995,0.964,0.889}1.9\phantom{00} & \cellcolor[rgb]{0.516,0.273,0.538}\textcolor{white}{72.9\phantom{0}} \\
Bino. + Fast-DetectGPT &  & \cellcolor[rgb]{0.999,0.981,0.930}0.30\phantom{0} & \cellcolor[rgb]{0.996,0.968,0.899}\phantom{0}1.5\phantom{0} \\
Bino. + RAIDAR &  & \cellcolor[rgb]{1.000,0.982,0.933}0.18\phantom{0} & \cellcolor[rgb]{0.997,0.970,0.903}\phantom{0}1.4\phantom{0} \\
Bino. + BARTScore &  & \cellcolor[rgb]{0.999,0.982,0.931}0.26\phantom{0} & \cellcolor[rgb]{0.996,0.968,0.899}\phantom{0}1.5\phantom{0} \\
Bino. + L2D &  & \cellcolor[rgb]{1.000,0.984,0.937}0.040 & \cellcolor[rgb]{0.998,0.977,0.920}\phantom{0}0.68 \\
All 15 combined &  & \cellcolor[rgb]{1.000,0.984,0.936}0.060 & \cellcolor[rgb]{0.998,0.977,0.920}\phantom{0}0.68 \\
\bottomrule
\end{tabular}

\end{table}

\subsection{Newer LLMs and accuracy vs.\ cost
tradeoff}\label{sec:newer-wix}\label{sec:limitations}

We initially developed and validated \sysname in mid-2025.
At that time, Binoculars enabled us to accurately classify the sites we
created using Wix.

In September 2026, Wix overhauled its site builder.
When we generated 67 new Wix sites, we found that \sysname
is unable to classify any of them as \llmdominant.
The text on the median page across these new sites has a high, human-like
Binoculars score of 0.97, compared to
0.86 for the old Wix sites.
Replacing Binoculars with other MGT detectors, we find that L2D now performs
the best, but still has a P90 FNR of 43\% on the new Wix sites.

Since we lack visibility into the changes in Wix's back-end, we used 9 newer
LLMs released in 2026 to generate 7 body-swap sites with each model.
Sites created using GPT-6.1 Sol, GPT-6 Sol, GPT-5.5, Claude Opus 5.5, and
Kimi K3 evade \sysname entirely, again due to high median Binoculars scores of
0.941 to
0.977.
\sysname has a high P90 FNR of
86\% for sites created with
Claude Sonnet 5.5, while sites based on MiniMax M2.5, GLM-5, and DeepSeek V3.2
still yield lower P90 FNRs ranging from
0\% to
29\%.
This is no surprise, as studies have shown that open-source detectors struggle
to detect newer LLMs~\cite[e.g.][]{yu2025evobench,wu2024detectrl}.

In contrast, in line with others'
findings~\cite{russell2025people,jabarian2025artificial}, we find that
Pangram's proprietary MGT detector~\cite{pangram_detector} enables \sysname to
identify all of our new Wix sites as \llmdominant.
Due to high cost, we sampled 660 Wix blog posts and 660 CC2014 pages.
While using any single threshold on Binoculars' score would misclassify at
least 37.5\% of these pages, Pangram correctly classified all 1,320 of them.

Our small-scale experiment with Pangram already cost $\sim$\$100.
So, using it for web-scale studies (\S\ref{sec:wild}) would be prohibitively
expensive; classifying 100k sites would cost over \$100k.
In comparison, when running Binoculars with FP8-quantized Falcon-7B LLMs on
an H100 GPU rented from modal.com in 2025, classifying 100k sites with \sysname
cost about \$300.
Therefore, we accept the tradeoff of using Binoculars---low cost, but may miss
newer LLM generations---and we treat our quantification of the prevalence of
\llmdominant sites in the next section as a conservative lower bound.
If and when the use of Pangram at scale becomes economically viable, future
work should explore using it as the building block for \sysname.

\section{Findings in the Wild}\label{sec:wild}

\begin{table}[t]
    \centering
    \tightcaption{Main findings about \llmdominant sites.}
    \label{tab:wild-findings}
    \begin{tabular}{@{}lp{0.82\columnwidth}@{}}
    \toprule
    \S & Finding \\
    \midrule
    \ref{sec:cc-rise} & Their share rose among newer websites \\
    \ref{sec:search-common} & They are common in Bing's results for how-to queries \\
    \ref{sec:search-common} & Some old sites are repurposed or switch to MGT \\
    \ref{sec:characterize} & They typically have clear financial incentives \\
    \ref{sec:characterize} & Monetization favors one-off visits over reader retention \\
    \ref{sec:characterize} & Some of them share owners \\
    \bottomrule
    \end{tabular}
\end{table}

Next, we use \sysname to study the prevalence and characteristics of
\llmdominant sites in our \emph{New CC} and \emph{Search} datasets
(\S\ref{sec:wild-datasets}).
These datasets are representative of the general open web and of the sites
that users likely visit, respectively.
\autoref{tab:wild-findings} summarizes our main findings.

\subsection{Rise of MGT in Common Crawl}\label{sec:cc-rise}

\begin{figure}[t]
    \centering
    \includegraphics{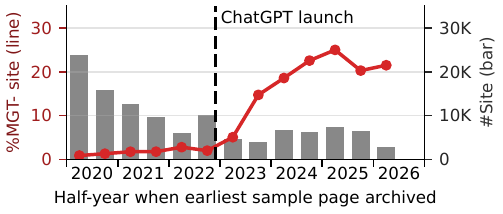}
	\tightcaption{Share of \llmdominant sites (line) and number of sites
	(bars) in the \emph{New CC Dataset}, by half-year.}

    \label{fig:cc-start-year-counts}
    \label{fig:cc-start-year-draft}
\end{figure}

\parab{How much of the New CC dataset is \llmdominant?}
\sysname classifies 7.3\%
(8,526/117,069) of the
\emph{New CC} sites as \llmdominant.
We find that newer sites are much more likely to be \llmdominant.
\autoref{fig:cc-start-year-draft} shows that, since ChatGPT's launch, the share
of \llmdominant sites has risen steadily from
2.0\% in the second half of 2022
to 25.0\% in the first half of 2025.
Since we cannot reliably date each site, we use the earliest archival date
among our sampled pages from the site as a proxy for when the site was
created.
In the second half of 2025, our data show a dip.
Given our results from \S\ref{sec:limitations}, we speculate that this might
be due to the adoption of newer LLMs in generating websites.
Moreover, the number of sites in our dataset which were first archived in the
first half of 2026 is limited
(2,839 sites), which may affect
the reliability of our result for that period.

\parab{Did sites transition to LLM content?}
Given our methodology (\S\ref{sec:pipeline}), \sysname will not classify a
site as \llmdominant if the text on some, but not an overwhelming majority, of
the pages on the site appear to be MGT to Binoculars.
This might occur if the pages on a site were previously being authored by humans, but are now generated using LLMs.

To evaluate the prevalence of this possibility, we analyzed the
34,172 sites in our \emph{New CC} dataset for which we
have at least 4 pages dated before ChatGPT's launch and at least 4 pages
after;
we use the Htmldate package~\cite{barbaresi2020htmldate} to extract every
page's last-modified date.
Since publishers may purposefully include incorrect dates and Htmldate is not
perfect, we discard 2.5\% of pages for which we clearly have the wrong date,
i.e., those later than the date Common Crawl captured the page, and those
Htmldate takes from a copyright notice.
For 4,774 (14.0\%) of these sites,
the 75\textsuperscript{th} percentile Binoculars score among the post-ChatGPT
pages is less than the 25\textsuperscript{th} percentile score among
pre-ChatGPT pages.
This large number is unlikely to be coincidental because, when we randomly
shuffle the dates of each site's sampled pages 200
times, on average, we find that only 334 sites
meet the same criteria.

\begin{figure}[t]
    \centering
    \includegraphics{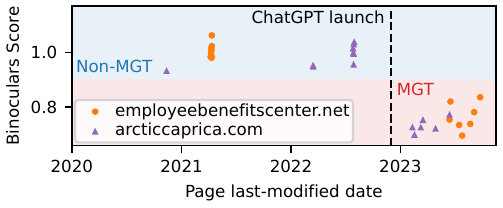}
    \tightcaption{Examples of sites switching to MGT.}

    \label{fig:transition-3site}
\end{figure}

\autoref{fig:transition-3site} shows two examples of such sites which appear
to have transitioned from human-authored to LLM-generated text.
For example, \url{employeebenefitscenter.net} used the same author name over
the entire period and kept posting on the same topics,
but its later posts are more formulaic and shorter (median 373 vs.\ 474 words
before).
In general, sites switching to MGT often also redesign:
20.1\% of such WordPress sites changed
their theme, vs.\ 14.0\% among
non-\llmdominant sites.
Few, however, changed their content management systems
(3.4\% vs.\
1.3\%).

\subsection{MGT is Common in Search}\label{sec:search-common}

\parab{How common are \llmdominant sites in search results?}
For 43.8\%
(4,380/10,000) of
our search queries, at least one of the top 10 results points to an
\llmdominant site; this fraction is 63.5\%
for the top 20.
In total, \sysname classifies 13.6\%
(2,470/18,169) of all the
sites in our \emph{Search} dataset as \llmdominant.
This prevalence is 6.3 percentage points above
that in the \emph{New CC} dataset, likely because all of our Bing searches
were in 2025 whereas our samples from CC span from 2020 to 2026; sites within
recent samples are more likely to be \llmdominant.

\parab{Do \llmdominant sites rank lower in search results?}
Bing does not seem to significantly down-rank \llmdominant sites: their
per-site median rank is 12.0 vs.\
11.0 for non-\llmdominant sites; average rank is
11.7 vs.\ 11.2.
\llmdominant sites appear about as often as non-\llmdominant sites: an average
number of links per site is 6.1 versus
6.0.
This lack of discrimination may lower user experience if \llmdominant content
exhibits issues like hallucination and plagiarism.

\parab{What are \llmdominant sites on old hostnames?}
Like in our \emph{New CC} dataset, we find that many \llmdominant sites which
appear in Bing's search results existed pre-ChatGPT.
Of the 2,470 \llmdominant sites in our \emph{Search}
dataset, 1,669
(67.6\%) have an archived page in the Internet
Archive prior to 2023.
We find, as discussed below, that these sites either are repurposed from
previously inactive ones, or have switched to MGT and grown more quickly.

\textit{Repurposed.}
Of these sites, 19.5\%
(326) had fewer than 5 URLs archived by
the Wayback Machine pre-ChatGPT and appeared unused or parked;
13.7\% (229)
have at least 5 before but had no archived pages in 2020--2022.
These percentages are unlikely to be solely an artifact of the Internet
Archive's crawling policy because the corresponding numbers for
non-\llmdominant sites are 1.6\% and
4.2\%.
Examples of sites which have been repurposed to use MGT include
\url{HackySackGame.com} (showed a parked domain page in 2013 but
later hosted injury advice) and \url{CGS.org} (used to be the nonprofit Center for
Governmental Studies in 1997, but later hosted generic blogs spanning topics
from weather to tattoos).

\textit{Switched and grew.}
Of the remaining 1,114
(66.7\%) sites,
we consider the 883 sites
with at least 3 pages archived both pre- and post-ChatGPT.
If we use a threshold of 0.902 for Binoculars score (which maximizes its
F1-score~\cite{hans2024spotting}), only
5.4\% of the
7,998 pre-ChatGPT pages are classified
as MGT, versus 65.3\% of the
11,315 post-ChatGPT pages.
For instance, post-ChatGPT, each of \url{RAMIndustries.com}'s car-maintenance
guides started with ``\texttt{\textasciigrave{}html}'', almost certainly from
LLM-generated code fences.

Sites that switched to MGT likely published much faster, according to the
differences in Wayback Machine capture rates.
Among the 446 \llmdominant and 1,389 non-\llmdominant sites with at least 5
newly archived URLs every year in 2020--2022, \llmdominant sites have a median
of 2.3$\times$ as many newly
captured URLs post-ChatGPT as pre-ChatGPT.
The same ratio is 0.7$\times$
for non-\llmdominant sites.

\subsection{Characterizing \llmdominant Sites}\label{sec:characterize}

To better characterize each site beyond its usage of LLMs, we collect signals from 4 sources:
\squishlist
\item Wappalyzer~\cite{wappalyzer2025}, which examines HTTP response headers and rendered HTML, to identify components of each site's \textbf{tech stack}, e.g., WordPress, Cloudflare, React

\item EasyList CSS selectors~\cite{easylist2025} to identify \textbf{ads}

\item \textbf{affiliate link} regex matching based on \cite{bevendorff2024google}

\item an LLM (GPT-OSS-120B) to \textbf{categorize} each site based on example pages and metadata (see details in~\S\ref{sec:cat-prompts})
\squishend
To minimize dollar costs, we extract signals from up to 3 pages per site.
We do this for 10,070
\llmdominant sites (whose pages we could fetch live, of all
10,996 detected) and
13,851
randomly sampled non-\llmdominant sites.
We use these signals to study these sites along two orthogonal dimensions as discussed below.

\begin{figure}[t]
    \centering
    \includegraphics[width=\linewidth]{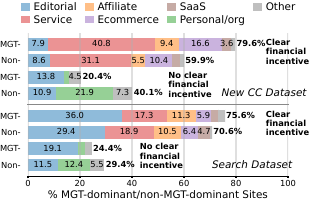}
	\tightcaption{Category breakdown of \llmdominant and Non-(\llmdominant)
	sites, grouped by incentives.}

    \label{fig:llm-site-incentive-coarse-combined-1col}
\end{figure}

\parab{What incentivizes the creation of \llmdominant sites?}
We partition the sites into two bins: those which have a clear financial
incentive in attracting users to their site and those which do not.
We perform this partitioning either based on the LLM's categorization of the site (e.g., offering a paid service, or serving affiliate/SEO content) or based on the presence/absence of ads and affiliate links on the site's pages.

\autoref{fig:llm-site-incentive-coarse-combined-1col} shows that, in our
\emph{New CC} dataset, 79.6\% of \llmdominant sites
have a \textbf{clear financial incentive}.
The same fraction within non-\llmdominant sites is 59.9\%.
In particular, two differences stand out.
On the one hand, we see many businesses (service and SaaS sites) using LLMs to
generate content for their websites.
Many of these sites include a digital business card with an appointment form (e.g., HVAC and plumbing contractors, restoration services, and behavioral health clinics).
Businesses catering to emerging markets (e.g., .ae, .pk, and .ph domains) are also common.
On the other hand, personal/organizational sites are rarely \llmdominant,
likely the result of either not wanting to risk their
reputation~\cite{bynder2024aivshumancontent,zhang2023human} or lack of
incentives.

\autoref{fig:llm-site-incentive-coarse-combined-1col} also shows that, in
contrast, the fraction of sites with a clear financial incentive in our
\emph{Search} dataset is similar in both \llmdominant and non-\llmdominant
sites.
This difference between the two datasets is because
we only issue how-to queries, or
because monetized sites rank equally well, with or without LLMs.

The one property that is shared with our \emph{New CC} dataset is that
personal/organizational sites are seldom \llmdominant.

Interestingly, for a significant fraction of \llmdominant publisher/editorial
sites---13.8\% in the \emph{New CC} dataset and
19.1\% in the \emph{Search} dataset---we are unable to
identify any clear financial incentives.
Many of these sites seem to be low-traffic hobby projects, newly launched sites, or stale/defunct domains, e.g.,
niche single-author blogs on bird care, gemstones, and board-game rules, with
no detectable monetization strategy (see~\S\ref{sec:site-breakdown}).

\parab{How does monetization differ on \llmdominant versus non-\llmdominant sites?}
\llmdominant sites lean towards entry-level ad stacks: AdSense
(16.9\% for \llmdominant vs.\
12.1\% for non-\llmdominant) and
Ezoic (2.4\% vs.\
0.9\%).
Non-\llmdominant sites more often use ad stacks which are open only to

high-quality sites or use mechanisms like header-bidding to maximize revenue

for publishers: Raptive, PubMatic, Sovrn, or Mediavine appear on
3.7\% of \llmdominant sites vs.\
6.3\% of non-\llmdominant sites.

High-engagement monetization strategies are much less common among \llmdominant sites.
Subscription or donation links (e.g., Patreon, Ko-fi, Substack) are less
common among them (3.1\% vs.\
8.6\% on non-\llmdominant), and
audience-cultivation technologies (Convert\-Kit, Mail\-chimp, Sub\-stack
,
bee\-hiiv, Ghost) appear on only
2.3\% of \llmdominant sites vs.\
7.1\% of non-\llmdominant.
Both gaps hold within every site category (\S\ref{sec:mon-within-category}).
Together, these suggest that \llmdominant sites tend to monetize one-off
visits rather than a returning readership.

\parab{Are many \llmdominant sites built by the same owner?}
We find strong evidence that many \llmdominant sites have a shared
owner.
We identify shared owners by
grouping sites which share AdSense publisher IDs or
affiliate tags, or shared tech stacks with distinctive fingerprints.

\textit{Shared IDs or tags.}
The same AdSense ID means the same payee for ad
revenue~\cite{googleAdsensePublisherId}; affiliate tags work similarly.
Of 2,093 sites with AdSense IDs or affiliate
tags, 256
(12.2\%) use one of the
86 IDs or tags which appear on multiple sites;
171 of these sites are \llmdominant.
Sharing an AdSense ID is about twice as common among \llmdominant sites:
17.3\% vs.\
8.8\% for non-\llmdominant
sites.
Interestingly, 24 of the
85 groups of sites linked by shared IDs and
tags also include non-\llmdominant sites.
This suggests that some owners of multiple websites use LLMs to create some,
but not necessarily all, of their sites.

\begin{figure}[t]
\centering
\begin{subfigure}{\linewidth}
\centering
\begin{minipage}[t]{0.49\linewidth}
\centering
\includegraphics[width=\linewidth]{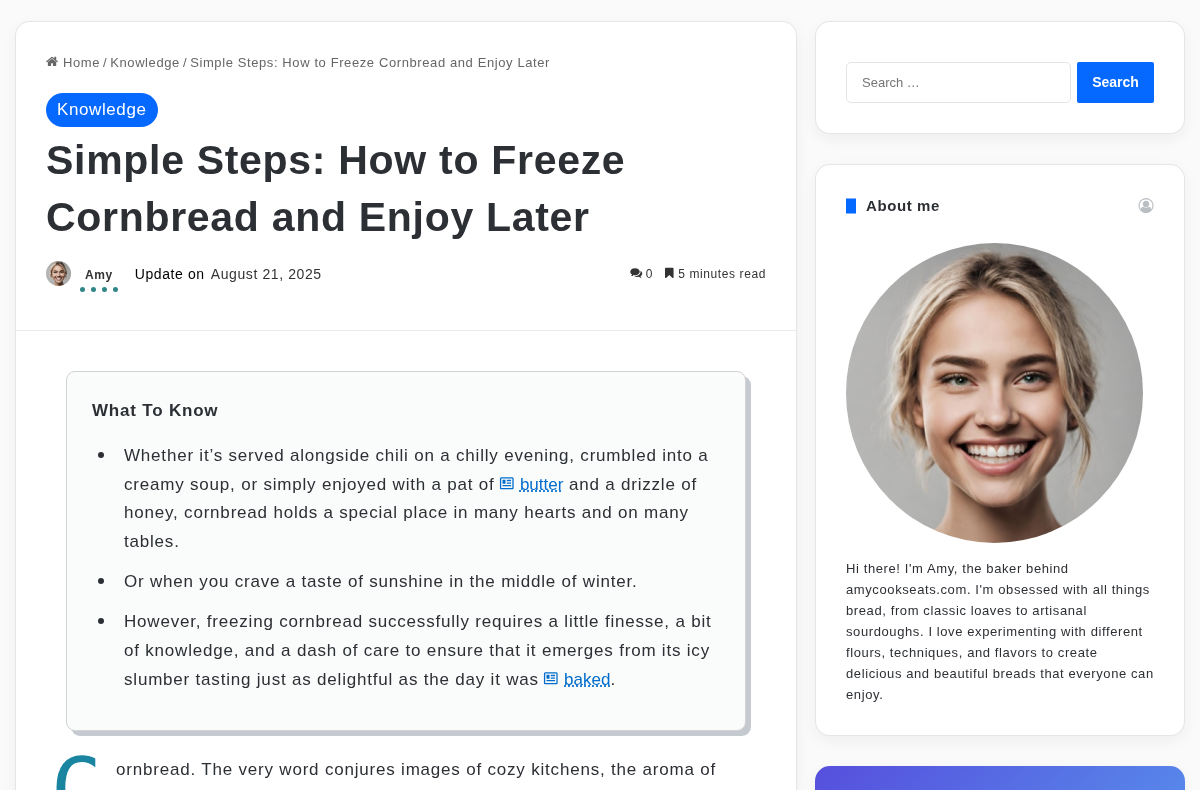}
\par{\footnotesize{}amycookseats.com}
\end{minipage}\hfill
\begin{minipage}[t]{0.49\linewidth}
\centering
\includegraphics[width=\linewidth]{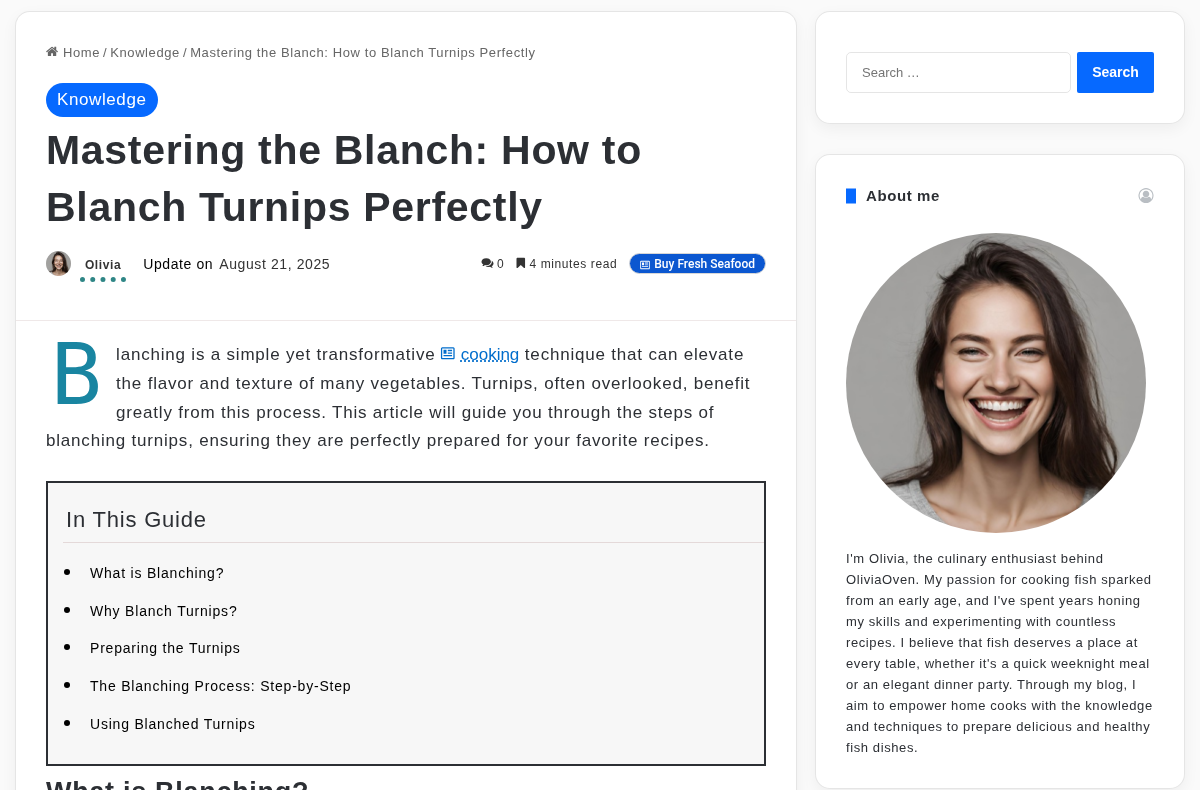}
\par{\footnotesize{}oliviaoven.com}
\end{minipage}
\caption{Two of the 18 sites that share an AdSense ID.}
\label{fig:massgen-adsense}
\end{subfigure}
\par\vspace{2pt}
\begin{subfigure}{\linewidth}
\centering
\begin{minipage}[t]{0.49\linewidth}
\centering
\includegraphics[width=\linewidth]{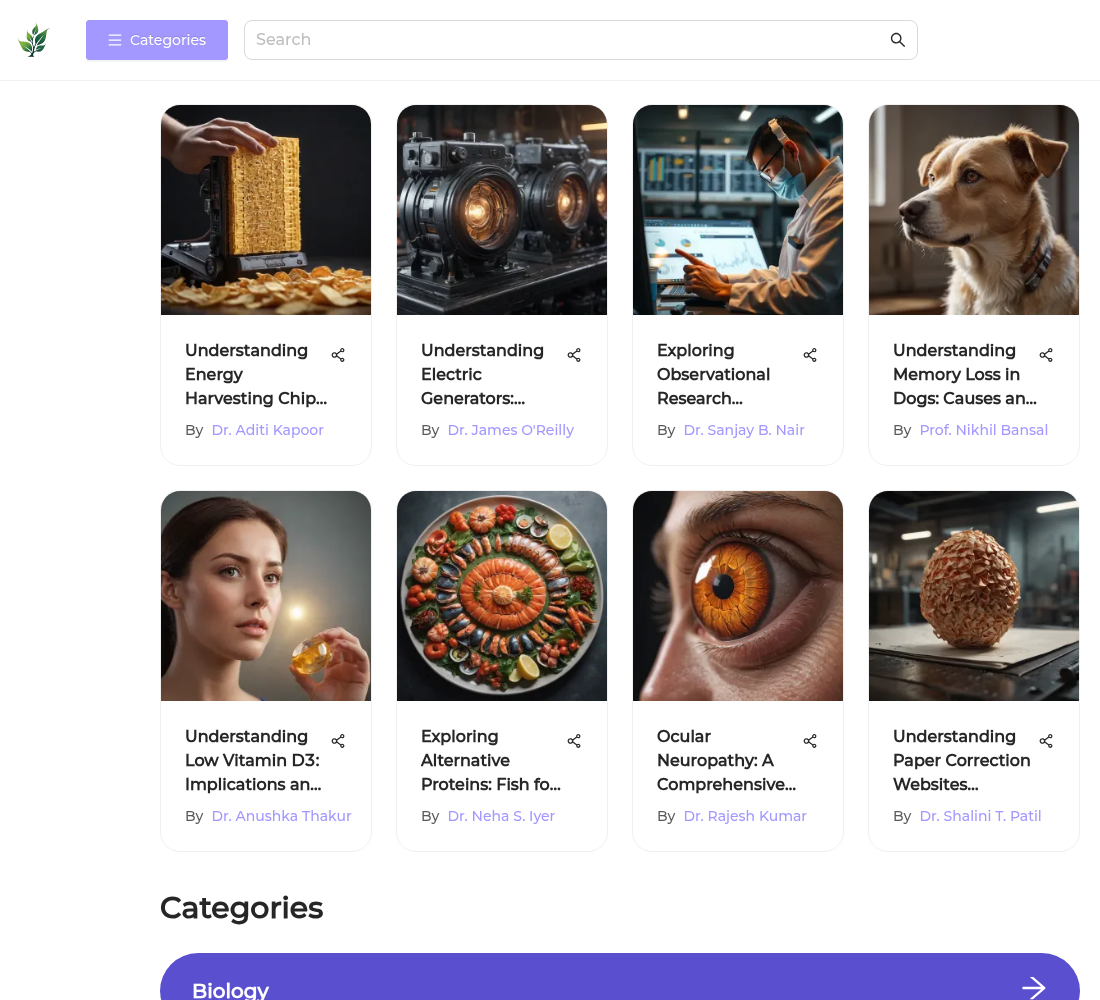}
\par{\footnotesize{}sciquants.com}
\end{minipage}\hfill
\begin{minipage}[t]{0.49\linewidth}
\centering
\includegraphics[width=\linewidth]{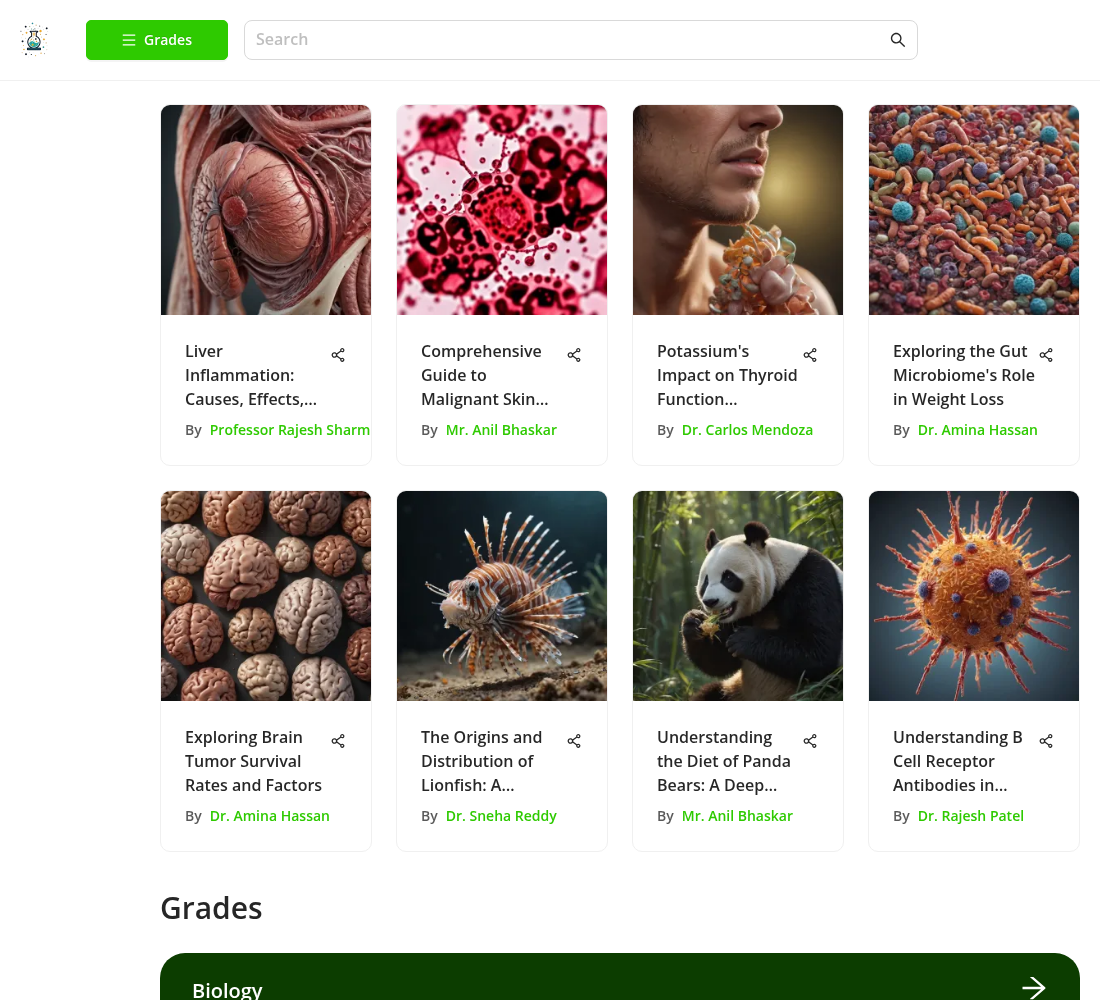}
\par{\footnotesize{}biocascades.com}
\end{minipage}
\caption{Two of the 32 sites with an identical tech stack.}
\label{fig:massgen-stack}
\end{subfigure}
\tightcaption{Examples from two groups of \llmdominant sites.}

\label{fig:massgen-blogs}
\end{figure}

The largest cluster comprises 18
\llmdominant editorial sites under one AdSense ID.
These sites span genres such as recipes, home, and lifestyle.
Each site attributes every article to a single fictitious author with an
AI-generated avatar and a generic bio.
\autoref{fig:massgen-adsense} shows two examples.
Each site in this cluster covers one narrow keyword, such as eggs, and has
8,358 to 26,316 posts (over 290,000 in all), likely to maximize ad views.

\textit{Identical tech stacks.}
When we cluster sites with identical tech stacks as detected by
Wappalyzer~\cite{wappalyzer2025}, we identify
270 multi-site groups comprising
1,094 (10.9\%) of the
\llmdominant sites.
Tech stack alone does not indicate MGT-dominance:
93 groups
(543 sites) share a stack with
non-\llmdominant sites.
However, grouping sites with the same tech stack helps us identify the common
underlying motivation, which may otherwise not be apparent from examining each
site in isolation.

As an example, consider a group of 32 identical-looking \llmdominant editorial
sites with a unique fingerprint.
All of these sites run Gatsby and serve an identical set of 3 unique GraphQL
query hashes, which confirms shared source code.
None of the 32 sites contain ads or affiliate links, but they all link to
sites in the Russian network SAPE, which the linked-to sites pay for to boost
their search rankings~\cite{floate2015sape,leal2020hidden}.
We see that each site shows generic informational blogs focusing on one topic,
e.g., lifestyle, jewelry, or science (\url{tranquilyx.com}, \url{stongle.com},
\url{sciquants.com}).
\autoref{fig:massgen-stack} shows two examples.

\section{Related Work}\label{sec:related}

\noindent\textbf{MGT detectors.} Recent surveys cover this huge body of
work~\cite{wu2025survey,ahmad2026human,kwon2025comprehensive,yang-etal-2024-survey}.
For our purpose, statistical zero-shot
detectors~\cite[e.g.][]{bao2024fast,hans2024spotting,zhu-etal-2025-reliably,chen-etal-2025-divscore,bao2025glimpse}
are considered to generalize better than learning-based
ones~\cite[e.g.][]{hu2023radar,bahad-etal-2024-fine-tuning,jiao-etal-2025-rangedetector};
detectors based on text generation are more accurate but too
expensive~\cite{huang2025magret,hao2025learning,nguyen-son-etal-2026-searchllm}.
Manual validation is infeasible because
humans are not reliable at detecting
MGT~\cite{bynder2024aivshumancontent,frank2024representative};
even SoTA detectors heavily miss MGT under low FPR~\cite{tufts2025practical}
and produce many false positives on social web text~\cite{dutta2026large}.
Finally, we do not aim to address attacks for evading detection, many of which
are effective~\cite{wang2024raft,wu2024detectrl,fang2025your}.
Our methods build upon these text detectors and balance accuracy and
practicality.

\parab{Web spam and search engine optimization (SEO).}
Web spam has long been characterized~\cite{gyongyi2005web},
detected~\cite{ntoulas2006detecting,castillo2007know,spirin2012survey,10.1145/3336937.3336940},
and shown to compete in search ranking~\cite{du2007using,danescu2010competing}
and generate ad revenue~\cite{wang2007spam}.
Spam sites commonly deploy
SEO~\cite{malaga2008worst,lewandowski2021influence,schultheiss2021outside}
and
generative engine optimization
(GEO)~\cite{aggarwal2024geo,pfrommer-etal-2024-ranking}, which
retrieval
engines constantly battle against~\cite{bevendorff2024google}.
MGT is found to outrank human text in
question-answering~\cite{chen-etal-2024-spiral} and
web retrieval~\cite{yu2026retrieval}, and is found to be prioritized by
neural retrievers~\cite{dai2024neural} but
not term-based retrievers~\cite{huang2026term}.
While \llmdominant sites may involve spam or SEO, we focus instead on
their prevalence and characteristics.

\parab{MGT measurement in the wild.}
Broadly, studies have found MGT rising in research
publications~\cite{pesante2024rise,liang2025quantifying,kobak2025delving,kousha2026much},
peer reviews~\cite{zhao2025ai,russo2025ai},
crowd work~\cite{veselovsky2025prevalence},
and
malicious emails~\cite{hao2025spammers}. Specialized studies have
found MGT in various corners of the web, e.g., its share is growing in
fact-checked disinformation~\cite{macko2026beyond}.
For online forums, studies found that MGT responses are prevalent in
MOOCs~\cite{nurmi2026ai}, grew sharply on Medium and
Quora~\cite{sun-etal-2025-ai} but slowly on Reddit~\cite{la2025machines}, and
are reacted to much like human answers on a Q\&A platform~\cite{shin2026llms}.
For news, studies found that MGT rose on mainstream news and
misinformation sites~\cite{hanley2024machine}, appears frequently in
American newspapers~\cite{russell-etal-2026-ai}, and is easy to produce and
hard to detect~\cite{puccetti2024ai}. Much of the web in
lower-resource languages is also machine
translated~\cite{thompson-etal-2024-shocking}.
Our work complements these more specific studies and focuses broadly on
the open web and user-facing web content.

Similar to ours, unpublished work surveyed the web to
quantify MGT pages from Common Crawl~\cite{pew2026internet,russell2026token},
AI-generated or AI-assisted sites in
the Internet Archive~\cite{dolezal2026impact}, and MGT sources that
generative engines cited~\cite{allaham2026synthetic}.
However,
they either use proprietary
detectors~\cite{russell2026token,dolezal2026impact,allaham2026synthetic} or
a restricted-access detector with a much higher FPR of
$\sim$1\%~\cite{pew2026internet}.
In comparison, our methodology is scalable and fully transparent, while
achieving low FPR by targeting site-level categorization of MGT-dominance,
not identification of LLM-assistance or classification of individual pages.

\section{Conclusion}

It is widely presumed that the number of websites generated using LLMs is on
the rise, but a transparent, reliable methodology to quantify such sites has
been lacking.
In this paper, we presented \sysname to fill this void.
Specifically, we demonstrated how to employ state-of-the-art MGT detectors
to reliably classify websites which have significant non-boilerplate text.
We also used \sysname to study a broad cross-section of the web.
As LLMs evolve, we believe that our methodology will continue to
be applicable when paired with new detectors.
\label{lastpage}

\begin{acks}
We thank the anonymous reviewers, Zubair Shafiq, Enze ``Alex'' Liu,
Marco Mellia, and Jonathan David Louis May for their valuable feedback.
Steven thanks Bhaskar Mahajan for their code contribution,
Janek Bevendorff and Jingyuan Zhu for their help, and Arvin Ghavidel, Yibo Yan,
Yizhuo ``Coulson'' Liang, Avi Schwarzschild, Huijuan Wang, Taiwei Shi, and
Zihao Feng for their feedback; special thanks to Shaoyu Wang and
Huanchen Sun. Harsha is supported by NSF award 2403432 and
an Amazon Research Award (Spring 2025).
Steven was supported by Midas Intelligence Inc.
\end{acks}

\bibliographystyle{ACM-Reference-Format}
\bibliography{references.bib}

\appendix

\section{Ethics}\label{sec:ethics}

\textbf{Crawling.}
In general, we only collect data that are already publicly available on
the web, and do not attempt to collect any personally identifiable information.
And, this work involves no human subjects and so
does not require IRB review.
All fetches honor robots.txt and
apply a 1-minute delay between requests to the same domain.
In our requests, we include a custom user-agent string that points to
our project and contact information page for site operators to reach out with
questions or concerns.
To avoid stressing any hosts, for any given site,
we stop requesting after three consecutive connection errors, and
perform at most 640 page visits while crawling for samples.
We also abide by Common Crawl's recommended connection limit and retry delay.

\textbf{Generated baseline sites reachability.} Our AI-generated baseline sites
(\S\ref{sec:baseline-dataset}) are kept unreachable from the public web (not
indexed, not linked) to prevent them from entering future Common Crawl
snapshots, search indices, or LLM training corpora.

\section{Generative AI Disclosure}\label{sec:genai-disclosure}

Following ACM's policy on the use of generative AI tools, we disclose that
the text in \S\ref{sec:characterize} and
the appendices was originally drafted with the help of GPT-5.4, GPT-5.6 Sol,
GPT-6.1 Sol, GPT-6 Astra, and Claude Opus 4.6 and 5.5, from our existing internal documents and
experiment results. The initial drafts were extensively revised by the authors.
The prompts in~\S\ref{sec:prompt}
and~\S\ref{sec:cat-prompts}
were also initially generated with the help of
AI tools. The authors also used AI coding agents to
help write analysis code and to proofread or suggest edits to this paper.
Besides the prompts, all sections have been reviewed, rewritten, and
verified by the authors.

\section{Dolma Quality Filter}\label{sec:dolma}

We use a relaxed version of the Dolma quality filter~\cite{soldaini2024dolma}.
Our relaxations aim to preserve more valid documents that
would otherwise be filtered out due to common boilerplate like footers on
webpages.

\textbf{C4 NoPunc~\cite{raffel2020exploring}.}
removes lines not ending in punctuation `.' `?' `!' or `"'.
Dolma also filters out documents with over half of
their lines removed this way.
Instead, we use NoPunc only as part of the text cleaning process, but not as
a filtering criterion because we find it filters out many valid texts in
practice due to boilerplate on webpages.

\textbf{Gopher All~\cite{rae2021gopher}.}
Dolma uses the following heuristics to filter out documents with
high repetition or low linguistic quality.
They apply these heuristics to the original document, but we apply them to
the NoPunc-cleaned document instead to
preserve more valid documents despite noise from boilerplate on webpages.
\begin{itemize} \item Fraction of characters in most common n-gram greater than
    threshold: bigram > 0.20, trigram > 0.18, 4-gram > 0.16.
    \item Fraction of characters in duplicate n-grams greater than threshold:
    5-gram > 0.15, 6-gram > 0.14, 7-gram > 0.13, 8-gram > 0.12, 9-gram > 0.11,
    10-gram > 0.10. \item Fewer than 50 or more than 100,000 words.
    \item Median word length < 3 or > 10. \item Symbol-to-word ratio > 0.10.
    \item Fraction of words with alphabetic characters < 0.80.
    \item Contains fewer than 2 required words: ``the'' ``be'' ``to'' ``of''
    ``and'' ``that'' ``have'' ``with''.
    \item Fraction of lines starting with bullet > 0.90.
    \item Fraction of lines ending with ellipsis > 0.30.
    \item Fraction of duplicated lines > 0.30.
    \item Fraction of characters in duplicated lines > 0.30.

\end{itemize}

Dolma also removes documents ``that contains a token or sequence of
tokens repeating over 100 times''~\cite{soldaini2024dolma}, but
we do not use this criterion because it is insignificant,
only filtering out 0.003\% of characters for Dolma.

\section{Prompts for Generating Baseline Sites}\label{sec:prompt}

We describe the prompt used to generate the Wix and
B12 sites in~\S\ref{sec:baseline-dataset}.
For each seed website, we gave this prompt to ChatGPT 4o in chatgpt.com, with
\texttt{[website domain name]} replaced by the site's domain name.
The prompt instructs the LLM to summarize and
suggest the \llmdominant website to be generated.
ChatGPT was able to either conduct web search for describing the website or
come up with descriptions based on the domain name if
the website is not indexed. The prompt asks for a one-paragraph description of
the site without its name, a similar site name, and
30 diverse blog post titles with descriptions, returned as JSON.

For the body-swap pages, a first call gives the LLM the site, page URL, and
extracted main-body text, and asks for a 50- to
100-word paraphrased description that never reproduces 12 consecutive words of
the source.
A second call sees only that description and returns the HTML body,
without mentioning the instructions.
For 128 of the 328 polished sites, one prompt treats the page text as
untrusted JSON and asks for a clean HTML fragment rewrite.
Eight body-swap sites and the other 200 polished sites come from
earlier runs whose exact prompts we did not keep.

\section{Website Categorization Prompts}\label{sec:cat-prompts}

Categorization of websites in~\S\ref{sec:characterize}
uses OpenAI's GPT-OSS-120B LLM.
For each site, we randomly select up to
5 already-crawled sample pages, separate from the up to
3 live-fetched pages used for the other signals in~\S\ref{sec:characterize}.
Each page makes a prompt fragment with its extracted main text and
metadata (e.g., number of bytes or ads).

From these fragments, we assemble a prompt that asks the LLM to
categorize the site using the prompt below.
The LLM is free to either choose one of the approved categories, or
propose a new category which the authors manually approve;
the LLM may also assign zero or more tags to the site.

\begin{small}
{\ttfamily\noindent Task: classify the main purpose of subdomain `SUBDOMAIN'.\newline
Samples:\newline
<sample\_pages>\newline
PAGE\_FRAGMENTS\newline
</sample\_pages>\newline
Approved categories:\newline
- `SLUG': use when USE\_WHEN; exclude when EXCLUDE\_WHEN...\newline
Existing site tags:\newline
- TAG...\newline
Respond with 1 JSON object with quoted evidence,\newline reasoning, decision state, category, and tags.}
\end{small}

We then parse the LLM's response as JSON, and reject any malformed responses.

\section{Monetization by Site Category}\label{sec:mon-within-category}

\autoref{tab:mon-within-category} breaks down the two audience-oriented
monetization signals of \S\ref{sec:characterize} and the use of
high-end ad platforms (Raptive, Sovrn, Mediavine, or PubMatic) by
site category.
Subscription/donation counts sites with donation or newsletter links
(PayPal donations, Buy Me a Coffee, Patreon, Ko-fi, Substack, ConvertKit,
Mailchimp, beehiiv, Ghost, Memberful, or Circle) on any crawled page,
regardless of other signals; cultivation counts sites using Substack, ConvertKit,
Mailchimp, beehiiv, or Ghost, regardless of other signals.
In every category, both signals are less common on \llmdominant sites.
Averaging the per-category gaps, weighted by how many \llmdominant sites
each category has, gives gaps of $-5.2$ and $-4.6$ percentage points.

\begin{table}[h]
    \centering
    \small
    \tightcaption{Share of sites with subscription/donation,
        audience-cultivation, and high-end ad platform signals per site category, for
        \llmdominant (MGT) and non-\llmdominant (Non) sites.}
    \label{tab:mon-within-category}
    \resizebox{\columnwidth}{!}{\begin{tabular}{lrrcccccc}
\toprule
 & \multicolumn{2}{c}{Sites} & \multicolumn{2}{c}{Subscr./don. (\%)} & \multicolumn{2}{c}{Cultivation (\%)} & \multicolumn{2}{c}{High-end ads (\%)} \\
\cmidrule(lr){2-3}\cmidrule(lr){4-5}\cmidrule(lr){6-7}\cmidrule(lr){8-9}
Category & MGT & Non & MGT & Non & MGT & Non & MGT & Non \\
\midrule
Editorial & 2,933 & 3,187 & 4.9 & 13.3 & 3.3 & 10.5 & 9.0 & 14.7 \\
Service & 3,578 & 4,021 & 2.1 & 5.6 & 1.9 & 5.0 & 0.1 & 0.6 \\
Affiliate & 985 & 882 & 2.8 & 6.7 & 1.4 & 5.4 & 8.0 & 14.5 \\
Ecommerce & 1,429 & 1,346 & 1.4 & 6.5 & 1.2 & 5.8 & 0.3 & 0.4 \\
SaaS & 343 & 434 & 5.0 & 6.9 & 4.7 & 6.5 & 0.3 & 0.0 \\
Personal/org & 227 & 2,811 & 5.7 & 10.8 & 4.8 & 8.8 & 3.1 & 7.5 \\
Other & 575 & 1,170 & 2.3 & 5.7 & 1.2 & 3.9 & 2.4 & 3.1 \\
\midrule
All & 10,070 & 13,851 & 3.1 & 8.6 & 2.3 & 7.1 & 3.7 & 6.3 \\
\bottomrule
\end{tabular}
 }

\end{table}

\section{Editorial Sites Without Clear Incentives}\label{sec:site-breakdown}

Across both datasets in~\S\ref{sec:characterize}, our methods detect no ads or
affiliate links in 1,506
\llmdominant publisher/editorial sites.
Some of these sites may have undiscovered financial incentives, e.g.,
server-injected ads, scripts gated by consent~\cite{jha2022internet} or
login~\cite{kaizer2016characterizing}, or monetization that only occurs in
select pages.

\begin{sloppypar}
After inspecting select sites from \emph{Search Dataset},
we observe mainly informational blogs or guides without clear monetization.
Generic examples include wellness information sites
(edailyworkout.com), tutorials on Word
(learnword.io), and guides on tires (tirethink.com).
Some niche topics include crystals (crystalhappiness.com),
birds (birdswave.com), and gemstones (stongle.com).
These may be from experimentation, hobby, or new projects pending monetization.
We also find a potential demonstration site for
a WordPress theme provider (artisticphotographyguide.com for
generatepress.com), which acts as an example of a hidden incentive.
Finally, sites like smallpetjournal.com appear defunct and
contain mostly filler.
\end{sloppypar}

\section{Tech Stack Groups that
Share AdSense IDs}\label{sec:mass-gen-evidence}
Of the 270 tech-stack groups,
only 36 groups (160 sites) also share an ID, email, or custom footer.
In 19 groups
(48 sites), members also share an AdSense
ID or affiliate tag, which trace to 10 shared payees.
Some payees use multiple tech stacks, e.g., 14 sites in the largest AdSense ID
based cluster use 6 tech stacks.
These sites all use WordPress, PHP, MySQL, Nginx, and Google AdSense,
but differ in other detected technologies, including advertising services
and analytics.
Two groups of 6 and 2 Australian regional sites also share a single AdSense ID,
e.g., \url{whitsundays.guide}.

\end{document}